\documentclass[pdflatex,amsmath,eqsecnum]{aastex6}

\usepackage{graphicx}% Include figure files
\usepackage{bm}% bold math

\newcommand{\be}{\begin{equation}}
\newcommand{\ee}{\end{equation}}
\newcommand{\ba}{\begin{eqnarray}}
\newcommand{\ea}{\end{eqnarray}}
\newcommand{\ban}{\begin{eqnarray*}}
\newcommand{\ean}{\end{eqnarray*}}

\usepackage{graphicx}% Include figure files
\usepackage{dcolumn}% Align table columns on decimal point
\usepackage{bm}% bold math
\usepackage{color}
\begin{document}

%% LaTeX will automatically break titles if they run longer than
%% one line. However, you may use \\ to force a line break if
%% you desire.

\preprint{RUP-20-31}
\preprint{KEK-TH-2269}
\preprint{KEK-Cosmo-0264}

\title{Spins of primordial black holes formed in the radiation-dominated phase of the universe: first-order effect}

%% Use \author, \affil, plus the \and command to format author and affiliation 
%% information.  If done correctly the peer review system will be able to
%% automatically put the author and affiliation information from the manuscript
%% and save the corresponding author the trouble of entering it by hand.
%%
%% The \affil should be used to document primary affiliations and the
%% \altaffil should be used for secondary affiliations, titles, or email.

%% Notice that each of these authors has alternate affiliations, which
%% are identified by the \altaffilmark after each name.  Specify alternate
%% affiliation information with \altaffiltext, with one command per each
%% affiliation.

%\altaffiltext{1}{harada@rikkyo.ac.jp}
%\altaffiltext{2}{The Graduate University for Advanced Studies
%(SOKENDAI), 1-1 Oho, Tsukuba, Ibaraki 305-0801, Japan}
%\altaffiltext{3}{Centre for Theoretical Physics, Jamia Millia Islamia, New Delhi 110025, India}

%%% Authors with the same affiliation can be grouped in a single
%%% \author and \affil call.
\author{Tomohiro Harada\altaffilmark{1}}
\author{Chul-Moon Yoo\altaffilmark{2}} 
\author{Kazunori Kohri\altaffilmark{3,4}} 
\author{Yasutaka Koga\altaffilmark{1}}
\and 
\author{Takeru Monobe\altaffilmark{1}}%
%
%
%%% Notice that each of these authors has alternate affiliations, which
%%% are identified by the \altaffilmark after each name.  Specify alternate
%%% affiliation information with \altaffiltext, with one command per each
%%% affiliation.
\affil{}
\email{harada@rikkyo.ac.jp}
\altaffiltext{1}{Department of Physics, Rikkyo University, Toshima,
Tokyo 171-8501, Japan}
\altaffiltext{2}{Gravity and Particle Cosmology Group,
Division of Particle and Astrophysical Science,
Graduate School of Science, Nagoya University, Nagoya 464-8602, Japan}
\altaffiltext{3}{Institute of Particle and Nuclear Studies, KEK,
1-1 Oho, Tsukuba, Ibaraki 305-0801, Japan}
\altaffiltext{4}{The Graduate University for Advanced Studies (SOKENDAI),
1-1 Oho, Tsukuba, Ibaraki 305-0801, Japan}
\shortauthors{Harada, T., et al.}
\shorttitle{Spins of primordial black holes}

%% Mark off the abstract in the ``abstract'' environment. 
\begin{abstract}

The standard deviation of the initial values of the nondimensional Kerr parameter $a_{*}$ 
of primordial black holes (PBHs) formed in the radiation-dominated phase of the universe is
estimated to the first order of perturbation
for the narrow power spectrum. 
Evaluating the angular momentum at turn around 
based on linearly extrapolated transfer functions and peak theory, 
we obtain the expression
$\sqrt{\langle a_{*}^{2} \rangle} \simeq 4.0\times 10^{-3}
 (M/M_{H})^{-1/3}\sqrt{1-\gamma^{2}}[1-0.072
 \log_{10}(\beta_{0}(M_{H})/(1.3\times 10^{-15}))]^{-1}$, where $M_{H}$,
 $\beta_{0}(M_{H})$, and $\gamma$ are
the mass within the Hubble horizon 
at the horizon entry of the overdense region, 
the fraction of the universe which collapsed to PBHs at the scale
 of $M_{H}$, and 
a quantity which characterizes the width of the power spectrum, respectively. 
This implies that for $M\simeq M_{H}$, the higher the probability of the PBH
 formation, the larger the standard deviation of the spins, while
PBHs of $M\ll M_{H}$ formed 
through near-critical collapse may have 
larger spins than those of $M\simeq M_{H}$.
In comparison to the previous estimate, 
the new estimate has the explicit dependence on the ratio $M/M_{\rm H}$
and no direct dependence on the current dark matter density.
On the other hand, it suggests that the first-order
 effect can be numerically comparable to the 
 second-order one.

\end{abstract}

%% Keywords should appear after the \end{abstract} command. 
%% See the online documentation for the full list of available subject
%% keywords and the rules for their use.

%% From the front matter, we move on to the body of the paper.
%% Sections are demarcated by \section and \subsection, respectively.
%% Observe the use of the LaTeX \label
%% command after the \subsection to give a symbolic KEY to the
%% subsection for cross-referencing in a \ref command.
%% You can use LaTeX's \ref and \label commands to keep track of
%% cross-references to sections, equations, tables, and figures.
%% That way, if you change the order of any elements, LaTeX will
%% automatically renumber them.

%% We recommend that authors also use the natbib \citep
%% and \citet commands to identify citations.  The citations are
%% tied to the reference list via symbolic KEYs. The KEY corresponds
%% to the KEY in the \bibitem in the reference list below. 

\section{Introduction}
Recently, primordial black holes (PBHs) have been intensively
investigated not only as a realistic candidate 
for dark matter~\citep{Carr:2009jm,Carr:2016drx,Carr:2017jsz,Carr:2020gox} 
but also as a possible origin of black holes
of tens of solar masses that source gravitational waves
detected by LIGO and 
Virgo~\citep{Nakamura:1997sm,Sasaki:2016jop,Bird:2016dcv,Clesse:2016vqa,Raidal:2017mfl}.
Various kinds of mechanisms generating PBHs have been proposed. 
Among them, we will focus on PBHs formed as a result of collapse of 
primordial cosmological 
perturbation. After inflation 
generates perturbations at super-horizon scales, 
the scales successively enter the Hubble horizon in the radiation-dominated phase
and the perturbations can collapse to form PBHs if the amplitude of the 
perturbation exceeds some threshold value. 
The threshold values have been studied in terms of $\bar{\delta}_{H}$, 
the density perturbation
averaged over the overdense region at horizon 
entry~\citep{Carr:1975qj,Polnarev:2006aa,Musco:2008hv,Harada:2013epa,Musco:2012au,Harada:2015yda}, 
although this is currently discussed in more sophisticated way based on the compaction function~\citep{Shibata:1999zs,Musco:2018rwt, Germani:2018jgr,Escriva:2019nsa,Escriva:2019phb} and peak 
theory~\citep{Yoo:2018kvb,Yoo:2020dkz}.
Roughly speaking, the mass of the PBH is given by the mass $M_{H}$ contained 
within the Hubble horizon at the time of the horizon entry $t$, where $M_{H}\sim
c^{3}t/G$, although for the near-critical case $\bar{\delta}_{H}\simeq 
\bar{\delta}_{H,th}$, the scaling law 
$M/M_{H}\propto (\bar{\delta}_{H}-\bar{\delta}_{H,th})^{\beta}$ with $\beta\simeq 0.36$ implies 
the formation of PBHs of $M\ll M_{H}$~\citep{Niemeyer:1999ak,Musco:2008hv,Musco:2012au}. 

Thanks to the uniqueness theorem, isolated 
stationary black holes in vacuum are perfectly 
characterized by two parameters, the mass $M$ and 
the spin angular momentum $\vec{S}$. Alternatively, 
we can use the nondimensional spin angular momentum 
$\vec{a}_{*}={\vec{S}c}/{GM^{2}}$.
The statistical distribution of 
the spins is a key probe into the origin of the 
black holes. In gravitational wave observation of binary black holes 
by LIGO and Virgo, 
the effective spin parameter $\chi_{\rm eff}$ can be measured. 
Up to now, the observed data for the most of the binary black holes 
have been consistent  with $\chi_{\rm eff}=0$~\citep{LIGOScientific:2018jsj},
although there are some exceptions~\citep{LIGOScientific:2020stg}.
 
PBHs may have changed their spins from their initial values.
PBHs have evaporated away through Hawking
radiation if their masses are smaller than $\sim 10^{15}$ g.
The spin of the black hole enhances Hawking radiation and deforms its spectrum. 
A spinning black hole decreases its nondimensional Kerr parameter
$a_{*}:=\sqrt{\vec{a}_{*}\cdot \vec{a}_{*}}$
through the Hawking radiation, while a black hole much more massive than
$\sim 10^{15}$ g does not significantly change $a_{*}$ 
through the Hawking 
radiation~\citep{Page:1976ki,Arbey:2019jmj,Dasgupta:2019cae}.
PBHs change their spins very little in the
radiation-dominated phase~\citep{Chiba:2017rvs}, 
while it is proposed that mass 
accretion could change the spin of black holes in 
some cosmological scenarios (e.g.~\citet{DeLuca:2020bjf}).  

In this paper, we investigate the initial values of the spins of
PBHs. Recently, this issue has been discussed by 
many authors from different points of
view~\citep{Chiba:2017rvs,Harada:2017fjm,DeLuca:2019buf,He:2019cdb,Mirbabayi:2019uph}. 
Among them, \citet{DeLuca:2019buf} apply
\citet{Heavens:1988}'s approach 
to the first-order effect of perturbation
and give a clear expression
$\sqrt{\langle a_{*}^{2} \rangle}\sim \Omega_{\rm
dm}\tilde{\sigma}_{H}\sqrt{1-\gamma^{2}}/\pi $, where $\Omega_{\rm dm}$,
$\tilde{\sigma}_{H}$, and $\gamma:=\langle k^{2}\rangle /\sqrt{\langle
k^{4}\rangle }$ are the
current ratio of the dark matter 
component to the critical density,  
the standard deviation of the density perturbation at
horizon entry of the inversed wave number, and 
a quantity which characterizes the width of the power spectrum.
In this paper, we apply the same approach to this issue but reach a
different result. This paper is organized as follows. In Section II, we
define the angular momentum and give its expression to the first order
of perturbation in the region which collapses to a PBH. In Section III,
we estimate the angular momentum 
at turn around under the assumption of the narrow spectrum. 
In Section IV, we estimate the nondimensional Kerr parameter of the PBH.
Section V is devoted to summary and discussion in particular in
comparison to the previous works.
We use the units in which $c=1$ in this paper.

\section{Angular momentum}
\subsection{Definition}
We follow \citet{DeLuca:2019buf} for the definition of angular momentum.
If the spacetime admits a Killing vector field $\phi_{i}^{a}$ 
which is tangent to a spacelike hypersurface and generates a spatial
rotation on it,  
the angular momentum $S_{i}(\Sigma)$ contained in the region $\Sigma$ on the
spacelike hypersurface can be defined as a conserved charge 
in terms of the integral on the boundary $\partial \Sigma$ as~\citep{Wald:1984rg}
\begin{equation}
 S_{i}(\Sigma):=\frac{1}{16\pi G}\int_{\partial \Sigma}\epsilon_{abcd}\nabla^{c}(\phi_{i})^{d}
 =-\frac{1}{8\pi G}\int_{\Sigma}R^{ab}n_{a}(\phi_{i})_{b}d\Sigma,
\label{eq:SRicci}
\end{equation}
where $n^{a}$ is the unit vector normal to $\Sigma$. 
Using the Einstein equation
$
 G_{ab}=8\pi T_{ab},
$
Eq.~(\ref{eq:SRicci}) transforms to 
\[
 S_{i}(\Sigma)=-\int_{\Sigma} T^{ab} n_{a}(\phi_{i})_{b}d\Sigma. 
\]
Let us use the 3+1 decomposition of the spacetime
\begin{equation}
 ds^{2}=-\alpha^{2}d\eta^{2}+a^{2}(t)\gamma_{ij}(dx^{i}+\beta^{i}d\eta)(dx^{j}+\beta^{j}d \eta).
\label{eq:3+1}
\end{equation}
We assume 
that the matter field is given by a single perfect fluid described by 
\begin{equation}
 T^{ab}=\rho u^{a}u^{b}+p (g^{ab}+u^{a}u^{b}),
\label{eq:perfectfluid}
\end{equation}
where $u^{a}$ is the four-velocity of the fluid element,
and that the background spacetime is given by a flat FLRW spacetime, in which 
the line element is written in the conformally flat form: 
\[
 ds^{2}=a^{2}(-d\eta^{2}+dx^{2}+dy^{2}+dz^{2}). 
\]
We can naturally define 
$\phi_{i}^{a}$ the generator of a spatial rotation with respect 
to the peak of the density perturbation at ${\bf x}={\bf x}_{pk}$ as 
\[
 (\phi_{i})^{a}=\epsilon_{ijk}(x-x_{pk})^{j}\delta^{kl}
\left(\frac{\partial }{\partial x^{l}}\right)^{a}. 
\]
To the first order of perturbation from the flat FLRW spacetime, 
we find 
\[
 S_{i}(\Sigma)=(1+w )a^{4}\rho_{b}\epsilon_{ijk}\int_{\Sigma}(x-x_{pk})^{j}(v-v_{pk})^{k}d^{3}x,  
\]
in the gauge with $\beta^{k}=0$, where $v^{i}:=u^{i}/u^{0}$ and we
have assumed the equation of state $p=w\rho$.
The region $\Sigma$ should be taken as a region which will collapse into a black hole. 
Although the determination of $\Sigma$ is a nontrivial task, following \citet{Heavens:1988,DeLuca:2019buf}, we assume 
\begin{equation}
 \Sigma=\left\{{\bf x}|\delta({\bf x})>f\delta_{pk}\right\}.
\label{eq:equidensityassumption}
\end{equation}

We truncate the Taylor-series expansion of $\delta$ around the peak at the second order as 
\[
 \delta \simeq \delta_{pk}+\frac{1}{2}\zeta_{ij}(x-x_{pk})^{i}(x-x_{pk})^{j},
\]
where 
\[
 \zeta_{ij}:=\left.\frac{\partial^{2} \delta}{\partial x^{i}\partial x^{j}}\right|_{{\bf x}={\bf x}_{pk}}.
\]
This truncation is justified provided that physical quantities do not
change so steeply within $\Sigma$. 
Adjusting $x$, $y$, and $z$ axes to the principal ones, 
we obtain
\begin{equation}
 \delta \simeq \delta_{pk}-\frac{1}{2}\sigma_{2}\sum_{i=1}^{3}\lambda_{i}
((x-x_{pk})^{i})^{2},
\label{eq:quadraticprincipal} 
\end{equation}
where $\sigma_{j}$ and $\lambda_{i}$ are defined in
Appendix~\ref{sec:peaktheory}. 
Equations~(\ref{eq:equidensityassumption}) 
and (\ref{eq:quadraticprincipal}) imply that $\Sigma$ is given by an ellipsoid 
with the three axes given by 
\[
 a_{i}^{2}=2\frac{\sigma_{0}}{\sigma_{2}}\frac{1-f}{\lambda_{i}}\nu,
\]
where we have defined $\nu:=\delta_{pk}/\sigma_{0}$.

Taking the truncated Taylor-series expansion of $v-v_{pk}$ at $x=x_{pk}$
\[
 v^{i}-v^{i}_{pk}\simeq v^{i}_{~j}(x-x_{pk})^{j},
\]
we find 
\begin{eqnarray}
 S_{i}(\Sigma)&\simeq &(1+w)a^{4}\rho_{b}\epsilon_{ijk}
v^{k}_{~l}\int_{\Sigma}(x-x_{pk})^{j}(x-x_{pk})^{l}d^{3}x \nonumber \\
 &= &(1+w)a^{4}\rho_{b}\epsilon_{ijk}v^{k}_{~l}J^{jl},
\label{eq:Si}
\end{eqnarray}
where 
\begin{eqnarray*}
v^{k}_{~l}&:=&\left.\frac{\partial v^{k}}{\partial x^{l}}\right|_{{\bf x}={\bf x}_{pk}}, \\
 J^{jl}&:=&\int_{\Sigma}(x-x_{pk})^{j}(x-x_{pk})^{l}d^{3}x=\frac{4\pi}{15}a_{1}a_{2}a_{3}
\mbox{diag}  (a_{1}^{2}, a_{2}^{2}, a_{3}^{2}).
\end{eqnarray*}

Here we concentrate on a growing mode of linear scalar
perturbation, which is briefly summarized in 
Appendix~\ref{sec:perturbationtheory}.
According to peak theory~\citep{Bardeen:1985tr,Heavens:1988}, which is 
briefly introduced in Appendix~\ref{sec:peaktheory},
the distribution of the nondiagonal components of 
$v_{ij}$ is independent from that of the trace-free part of $J^{jl}$.
Then, we obtain
\begin{equation}
\sqrt{\langle S_{i}S^{i}\rangle }=S_{{\rm ref}}\sqrt{\langle s_{e}^{i}s_{ei}\rangle},
\label{eq:Sseparate}
\end{equation}
where 
\begin{eqnarray}
 S_{{\rm ref}}(\eta)&=&(1+w)a^{4}\rho_{b}g(\eta) (1-f)^{5/2}R_{*}^{5}, 
\label{eq:Sref}\\
 \vec{s}_{e}&=&\frac{16\sqrt{2}\pi}{135\sqrt{3}}\left(\frac{\nu}{\gamma}\right)^{5/2}
\frac{1}{\sqrt{\Lambda}}(-\alpha_{1}\tilde{v}_{23},\alpha_{2}\tilde{v}_{13},-\alpha_{3}\tilde{v}_{12}), 
\label{eq:se}
\\
 \alpha_{1}&=&\frac{1}{\lambda_{3}}-\frac{1}{\lambda_{2}},~~
 \alpha_{2}=\frac{1}{\lambda_{3}}-\frac{1}{\lambda_{1}},~~
 \alpha_{3}=\frac{1}{\lambda_{2}}-\frac{1}{\lambda_{1}},~~ 
 \Lambda:=\lambda_{1}\lambda_{2}\lambda_{3}, 
\end{eqnarray}
and $R_{*}$ and $\gamma$ are defined in Appendix~\ref{sec:peaktheory}.
The quantity $\gamma$ must satisfy $0\le \gamma\le 1$ and we can usually assume
$0.8\alt \gamma \le 1$ for PBH formation~\citep{DeLuca:2019buf}.
The function $g(\eta)$ is defined by 
\begin{equation}
\langle ({v^{k}_{~l}}(\eta))^{2} \rangle =g^{2}(\eta)\langle (\tilde{v}^{k}_{~l})^{2}\rangle,
\label{eq:decouple}
\end{equation}
for all $(k,l)$, where $\tilde{v}^{k}_{~l}$ is time-independent and  
defined in Eq.~(\ref{eq:vijtilde}). \footnote{In \citet{Heavens:1988,DeLuca:2019buf}, 
the condition $v^{k}_{~l}(\eta)=g(\eta)\tilde{v}^{k}_{~l}$ is assumed.}

\subsection{Long-wavelength solutions and near-spherical approximation}

Motivated by inflationary cosmology, 
we consider cosmological long-wavelength solutions as initial data 
at $\eta=\eta_{\rm init}$, 
in which the density
perturbation in the constant mean curvature (CMC) slicing is written in 
terms of the curvature perturbation $\zeta$ in the uniform-density slicing
as follows~\citep{Harada:2015yda}:
\begin{equation}
 \delta_{\rm CMC}=-\frac{1}{2\pi a^{2}\rho_{b}}e^{5\zeta/2}\Delta e^{-\zeta/2},
\label{eq:CLWLS}
\end{equation}
where $\Delta:=\delta^{ij}\partial_{i}\partial_{j}$ and $\zeta$ is defined as $\gamma_{ij}=e^{-2\zeta}\delta_{ij}$ in the
uniform-density slicing. We assume that the density perturbation is
appropriately smoothed at scales smaller than the one under 
consideration. 
(See e.g.~\citet{Yoo:2018kvb,Young:2019osy,Yoo:2020dkz,Tokeshi:2020tjq} for the possible dependence on the choice of the window functions.)

To make the situation clear, we will apply peak theory to this density
perturbation field.
For $\nu\gg 1$, peak theory implies that the density perturbation is
nearly spherical near the peak with 
$
 \lambda_{i}=({\gamma\nu}/{3})(1+\epsilon_{i})
$~\citep{Bardeen:1985tr, Heavens:1988}, 
where
$\epsilon_{i}=O ({1}/(\gamma\nu))$,
and hence we obtain
\begin{equation}
 a_{i}\simeq r_{f}=\sqrt{6(1-f)}\frac{\sigma_{0}}{\sigma_{1}}.
\label{eq:rf}
\end{equation}
That is, the region $\Sigma$ is nearly spherical and the 
deviation appears at the order of $1/\nu$.

In the following, we assume $w=1/3$ and without loss of generality 
take ${\bf x}_{pk}=0$.
Linearizing Eq.~(\ref{eq:CLWLS}), we obtain
\begin{equation}
 \delta_{\rm CMC}=\frac{2}{3 a^{2}H_{b}^{2}}\Delta \zeta.
\label{eq:LCLWLS}
\end{equation}
Therefore, we find
\[
 \sigma_{j}^{2}=\frac{4}{9}\eta^{4}_{\rm init}
\int\frac{dk}{k}k^{4+2j}P_{\zeta}(k),
\]
where we have assumed that $\zeta_{{\bf k}}(0)$
obeys a homogeneous Gaussian distribution with
\[
 \langle \zeta_{{\bf k}}(0)\zeta^{*}_{{\bf k}'}(0)\rangle
  =(2\pi)^{3}\delta^{3}({\bf k}-{\bf k}')|\zeta_{{\bf k}}(0)|^{2}
\]
and the power spectrum $P_{\zeta}(k)$ is defined as
$P_{\zeta}(k):={k^{3}}|\zeta_{{\bf k}}(0)|^{2}/(2\pi^{2})$.
\color{black}

As for the velocity gradient field, 
from Eq.~(\ref{eq:v}), 
we have
\begin{eqnarray}
{v}^{i}_{~j}(\eta, {\bf x}):=\left(\frac{\partial v^{i}}{\partial x^{j}}\right)(\eta,{\bf x})=
\int \frac{d^{3}{\bf k}}{(2\pi)^{3}}\frac{k^{i}k_{j}}{k}v_{{\bf
k}}(\eta)e^{i {\bf k}\cdot {\bf x}}.
\label{eq:vijbar}
\end{eqnarray}
Therefore, we obtain the following expression for $g(\eta)$ 
\begin{equation}
g^{2}(\eta)=\frac{4}{9}\int\frac{dk}{k}k^{2}T_{v}^{2}(k,\eta)P_{\zeta}(k),
\label{eq:getageneral}
\end{equation}
where $T_{v}(k,\eta)$ is a transfer function for $v_{{\bf k}}(\eta)$ 
defined in Appendix \ref{sec:perturbationtheory}
and we have used 
$\langle \tilde{v}^{i}_{~j}\tilde{v}^{j}_{~i}\rangle =1$
as seen in Eq.~(\ref{eq:variancetildevij}).

\section{Estimate of the angular momentum}

\subsection{Narrow power spectrum}
In general, we cannot expect a simple expression for $g(\eta)$ because 
it is obtained by a convolution of different modes with different time dependence.
In \citet{Heavens:1988}, this is possible because the growth rate function
is homogeneous at sub-horizon scales
in the Einstein-de Sitter universe.
In \citet{DeLuca:2019buf}, they implicitly assume that the perturbation of some single 
$k$ effectively determines the angular momentum of the region $\Sigma$.
Here, we assume the same assumption as in \citet{DeLuca:2019buf}.
This is possible if we assume the power spectrum 
has a narrow peak at $k=k_{0}$ so that 
\[
P_{\zeta}(k)\simeq \sigma_{\zeta}^{2} k_{0}\delta(k-k_{0}).
\]
Then, Eq.~(\ref{eq:LCLWLS}) implies
\begin{equation}
\sigma_{j}\simeq \frac{2}{3}\eta^{2}_{\rm init}k_{0}^{2+j}\sigma_{\zeta}
\label{eq:sigmajDiracdelta}
\end{equation}
and therefore $\gamma\simeq 1$. 
In this case, from Eq.~(\ref{eq:getageneral}), 
we can obtain
\begin{equation}
g(\eta)\simeq \frac{2}{3}k_{0}|T_{v}(k_{0},\eta)|\sigma_{\zeta}.
\label{eq:getanarrow}
\end{equation}
In more general case, $k_{0}$ is identified with $k$ which dominates the 
integral on the right hand side of Eq.~(\ref{eq:getageneral}).

According to peak theory, in the case of the narrow power spectrum, 
the most probable profile is given by a sinc function
\citep{Bardeen:1985tr,Yoo:2018kvb}, that is,
\begin{eqnarray*}
 \delta_{\rm CMC}(\eta,{\bf r})= \delta_{pk}(\eta)\psi(r),~~
\zeta (\eta,{\bf r})= \zeta_{pk}(\eta)\psi(r),~~
\psi(r)=\frac{\sin(k_{0}r)}{k_{0}r}.
\end{eqnarray*} 
Then, we can replace the harmonic function $Y$ with $\psi(r)$.
We identify $\delta_{pk}(\eta)$ 
with $\delta_{{\rm CMC},k_{0}}(\eta)$ in Eq.~(\ref{eq:deltaCMC}).
From Eq.~(\ref{eq:LCLWLS}), we obtain 
\begin{eqnarray}
\delta_{{\rm CMC}, k_{0}}(\eta)\simeq 
\frac{2}{3}x^{2}(-\zeta_{k_{0}}(0)),\quad 
D=\frac{4\sqrt{3}}{3}(-\zeta_{k_{0}}(0)),
\label{eq:D1deltaHpk}
\end{eqnarray}
where $x=k_{0}\eta$ and $D$ is defined in Appendix~\ref{sec:perturbationtheory}.

\subsection{PBH formation threshold}
Under the truncated Taylor-series expansion, 
since the initial density perturbation profile is given by 
\[
 \delta_{{\rm CMC},k_{0}}(\eta_{\rm init},r)\simeq \delta_{{\rm CMC},
  k_{0}}(\eta_{\rm init})
\left[1-\frac{1}{6}(k_{0}r)^{2}\right], 
\]
the compaction function $C_{\rm CMC}(\eta,r)$ 
in the CMC slicing is given in the
long-wavelength limit by 
\[ 
 C_{\rm CMC}(\eta_{\rm init},r):=\left(\frac{\delta
				  M}{ar}\right)(\eta_{\rm init},r)\simeq \frac{1}{3}(k_{0}r)^{2}\left[1-\frac{1}{10}(k_{0}r)^{2}\right](-\zeta_{k_{0}}(0)), 
\]
where $\delta M$ is the mass excess. This
is independent from $\eta_{\rm init}$. It takes a maximum value
$C_{\rm max}$ at $r=r_{m}$, where 
\[
 C_{\rm max}=\frac{5}{6}(-\zeta_{k_{0}}(0)),\quad r_{m}=\sqrt{5}k_{0}^{-1}.
\]
The threshold value of $C_{\rm max}$ for the PBH formation is known to
$C_{\rm max}\simeq 0.38-0.42 \simeq 2/5$ from fully nonlinear numerical
simulations and this is fairly stable against different 
profiles of Gaussian-function or sinc-function shape~\citep{Shibata:1999zs,Harada:2015yda,Musco:2018rwt,Germani:2018jgr}. Using the threshold value
$C_{\rm max}\simeq 2/5$, 
we can identify the threshold values for other variables as 
$
 \zeta_{k_{0}}(0)\simeq -{12}/{25}
$
or 
$
D\simeq 16\sqrt{3}/25.
$
For this value of $\zeta_{k_{0}}(0)$, we can calculate the density
perturbation $\bar{\delta}_{H}$ averaged over the overdense region with the
radius $r_{0}=\sqrt{6}k_{0}^{-1}$ in the long-wavelength limit 
at horizon entry $\eta_{\rm init}=\eta_{H}$, 
which we define $(aH)(\eta_{H}) r_{0}=1$. The result is the following:
\[
 \bar{\delta}_{H}= \bar{\delta} (aHr_{0})^{2} \simeq \frac{2}{5}\cdot 
\frac{2}{3}(k_{0}r_{0})^{2}(-\zeta_{k_{0}}(0))\simeq \frac{96}{125}=0.768.
\]
This is fairly consistent with the numerical value $\simeq 0.63-0.84 $ in the CMC slicing, 
which is obtained by converting the threshold value $\simeq 0.42-0.56$ in the comoving slicing  obtained in fully nonlinear numerical 
simulations for the Gaussian-function or sinc-function shaped profiles~\citep{Polnarev:2006aa,Musco:2012au,Harada:2015yda,Musco:2018rwt,Germani:2018jgr}.
Although the long-wavelength limit is only approximately 
valid at $\eta=\eta_{H}$, 
it is useful and conventional 
to use $\bar{\delta}_{H}$ obtained in 
the extrapolation of the long-wavelength limit to
$\eta_{\rm init}=\eta_{H}$ to refer to the 
amplitude of the density perturbation.
Alternatively, one can use the curvature perturbation 
$\zeta_{k_{0}}(0)$ for which the threshold and standard deviation are
given by $\simeq -12/25$ and $\sigma_{\zeta}$, respectively.
From now on, if we set 
$\eta_{\rm init}=\eta_{H}$ in Eq.~(\ref{eq:sigmajDiracdelta}), we denote
$\sigma_{0}$ with $\sigma_{H}$. 
Then, we find 
\[
 \sigma_{H}=4\sigma_{\zeta}.
\]
Using this notation, from Eq.~(\ref{eq:getanarrow}), we find
\begin{equation}
 g(\eta) \simeq \frac{1}{6}|T_{v}(k_{0},\eta)|k_{0}\sigma_{H}.
\label{eq:getaTvksigma}
\end{equation}
We should note that since $\bar{\delta}$, the density perturbation
averaged over the overdense region, is given by 
$\bar{\delta}\simeq (2/5)\delta_{pk}$, we find
\[
 \nu=\frac{\delta_{pk}}{\sigma_{0}}\simeq 
\frac{5}{2}\frac{\bar{\delta}_{H}}{\sigma_{H}}=\frac{5}{2}\bar{\nu},
\]
where we have defined $\bar{\nu}$ as $\bar{\nu}=\bar{\delta}_H/\sigma_{H}$.

\subsection{Decoupling from the cosmological expansion}
In \citet{DeLuca:2019buf},
the angular momentum of the region $\Sigma$ at the horizon entry of the
inversed wave number is identified 
with the initial spin angular momentum of the black hole 
by arguing that turn around is immediately after the horizon entry.
This might result in misestimating the nondimensional spin parameter
because the angular momentum increases and the mass decreases in time
during the cosmological expansion.
Here, we will estimate the angular momentum of the black hole by that 
of the region $\Sigma$
at turn around, after which the evolution of the region decouples from the 
cosmological expansion and the mass and the angular momentum of the collapsing region should be approximately conserved. 
However, it is not a trivial task to determine this moment.
Strictly speaking, turn around is beyond the regime of linear perturbation.
However, since it can be regarded as still in a quasi-linear regime, 
we should be able to apply an extrapolation of linear perturbation theory.
We here identify the condition 
$\delta_{\rm CMC}\simeq 1$ in the CMC slicing 
as the decoupling condition because this
implies that the local density perturbation becomes so large that 
the expansion should be about to turn around.

To go beyond the turn around, 
the CMC slicing should not be appropriate
because the maximum expansion 
means a vanishing mean curvature, while there exists a far region, where 
the mean curvature is nonvanishing due to Hubble expansion.
To avoid this difficulty, we will shift to the conformal Newtonian gauge. 
It is expected that the dynamics should fit to an usual Newtonian
picture later. For this reason, we evaluate $T_{v}(\eta_{\rm ta})$ in
Eq.~(\ref{eq:getageneral}) for $v_{\rm CN}$, the velocity perturbation
in the conformal Newtonian gauge at the decoupling from the cosmological
expansion.

\begin{figure}[htbp]
\begin{center}
\begin{tabular}{cc}
 \includegraphics[width=0.45\textwidth]{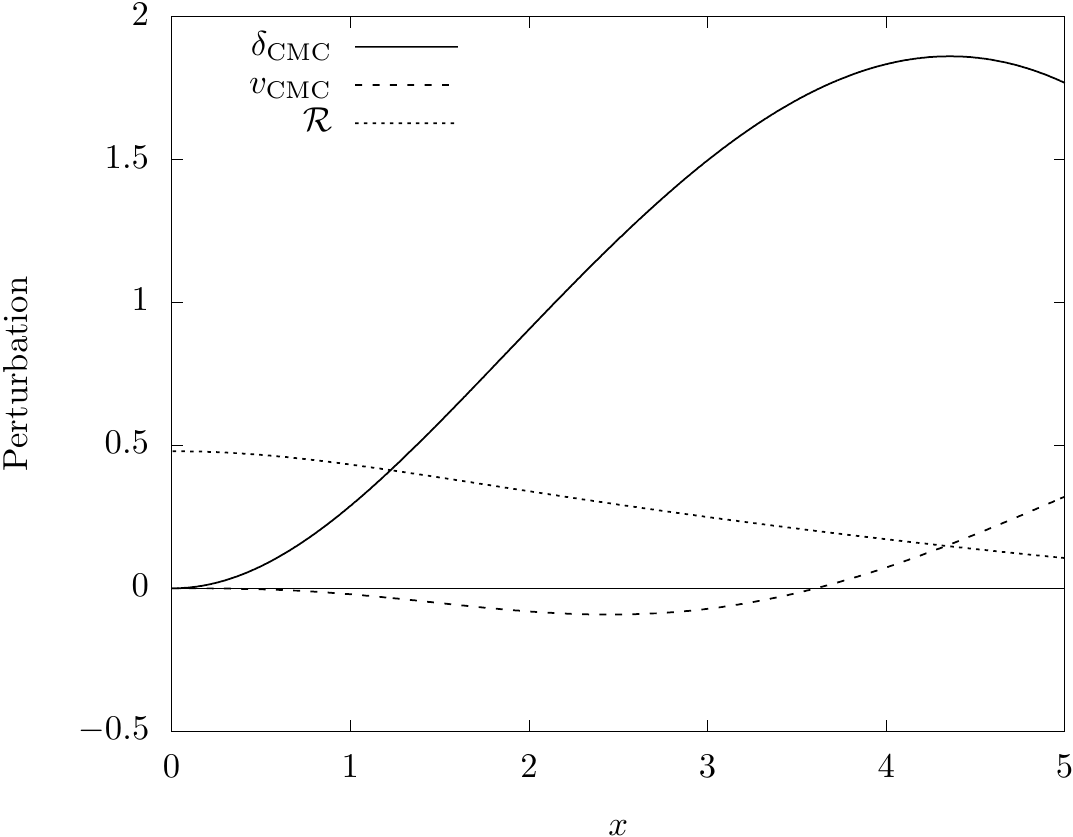}\label{fg:v_delta_R_cmc}
 &  
 \includegraphics[width=0.45\textwidth]{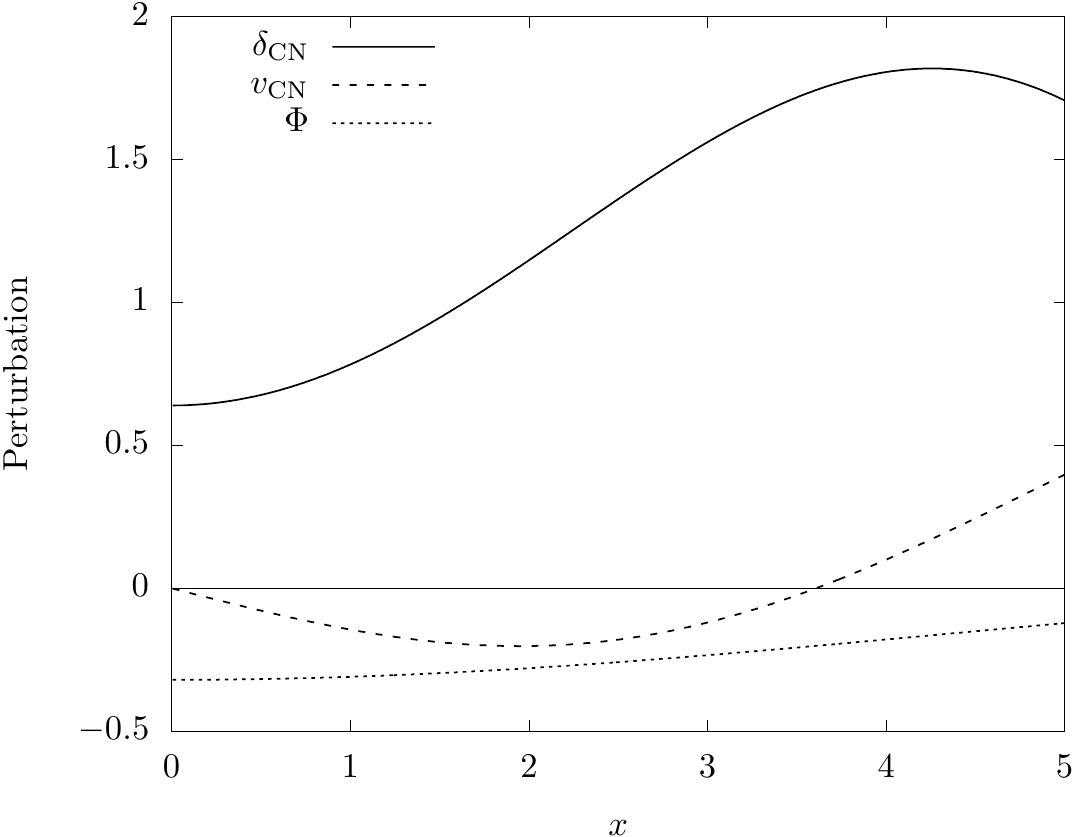}\label{fg:v_delta_phi_newton}
\end{tabular}
\caption{\label{fg:perturbations} 
The functions $v_{\rm CMC}$, $\delta_{\rm CMC}$, 
and ${\cal R}$ in the CMC slicing and 
$\delta_{\rm CN}$, $v_{\rm CN}$, and $\Phi$  
in the conformal Newtonian gauge are 
 plotted for $D= 16\sqrt{3}/25$ as functions of $x=k\eta$. 
We can see that $\delta_{\rm CMC}$ gets larger than unity at $x\simeq 2.14$,
 at which $v_{\rm CN}\simeq -0.199$.}
\end{center}
\end{figure}

In Fig. \ref{fg:perturbations}, we can see that the
turn around occurs $x=x_{\rm ta}\simeq 2.14$ 
for $D=16\sqrt{3}/25$. 
The value of the transfer function at the turn around $x=x_{\rm ta}$ is
calculated to give 
\[
T_{v_{\rm CN}}(k_{0},\eta_{\rm ta})=\frac{v_{\rm CN}(x_{\rm ta})}{\Phi_{
 k_{0}}(0)}\simeq 0.622, 
\]
where we have used $v_{{\rm CN}}(x_{\rm ta})\simeq -0.199$.
Thus, from Eq.~(\ref{eq:getaTvksigma}), 
the value of $g_{\rm CN}(\eta_{\rm ta})$ is given by  
\[
 g_{\rm CN}(\eta_{\rm ta})\simeq 0.104 k_{0}\sigma_{H}.
\]

Although there is some ambiguity in the choice of the decoupling condition and 
the gauge condition, it will not change the estimate 
in orders of magnitude as seen from Fig.~\ref{fg:perturbations} if we
choose $x_{\rm ta}$ between $\simeq 1.5$ and $\simeq 3$.

\section{Estimate of the nondimensional Kerr parameter}

\subsection{Estimate of $A_{{\rm ref}}$}

Let us estimate the reference spin value at turn around
\begin{equation}
 A_{{\rm ref}}(\eta_{\rm ta})=\frac{S_{{\rm ref}}(\eta_{\rm ta})}{GM_{\rm ta}^{2}}=\frac{\frac{4}{3}\left[a^{4}\rho_{b}g_{\rm CN} \right](\eta_{\rm ta})(1-f)^{5/2}R_{*}^{5}}{GM_{\rm ta}^{2}},
\label{eq:Aref}
\end{equation}
where the black hole mass $M$ is identified with the mass within the region $\Sigma$ at turn around
\[
 M_{\rm ta}=(\rho_{b}a^{3})(\eta_{\rm ta})\cdot \frac{4}{3}\pi r_{f}^{3}.
\]
This is different from $M_{H}$, which we define 
the mass within the horizon at the horizon entry of the overdense region.
The condition for the horizon entry $H^{-1}(\eta_{H})=ar_{0}$ 
implies $\eta_{H}=r_{0}$ or $x=\sqrt{6}$.
Since $a(\eta)\propto \eta$, we have 
\[
 \frac{a(\eta_{\rm ta})}{a(\eta_{H})}=\frac{\eta_{\rm ta}}{r_{0}}=\frac{x_{\rm ta}}{\sqrt{6}}.
\]
Using $\rho_{b}a^{3}\propto a^{-1}$, we find 
\begin{eqnarray*}
 M_{\rm ta} \simeq \frac{\sqrt{6}}{x_{\rm ta}}(1-f)^{3/2}M_{H}.
\end{eqnarray*}
Using Eq.~(\ref{eq:getaTvksigma}) and 
$2GM_{H}=a(\eta_{H})r_{0} $, we obtain a simple expression
\begin{equation}
A_{{\rm ref}}(\eta_{\rm ta})\simeq 
\frac{1}{24\sqrt{3}\pi} x_{\rm ta}^{2}(1-f)^{-1/2}|T_{v_{\rm CN}}(k_{0},\eta_{\rm ta})|\sigma_{H}.
\label{eq:newAref}
\end{equation}

\subsection{Estimate of $a_{*}$}

As for the distribution of $s_{e}$, we just quote 
the result of \citet{Heavens:1988} 
with the correction by \citet{DeLuca:2019buf}. For large $\nu$ limit, if we define $h$ by
\[
 s_{e}:=\sqrt{\vec{s_{e}}\cdot \vec{s_{e}}}= \frac{2^{9/2}\pi}{5\gamma^{6}\nu}\sqrt{1-\gamma^{2}}h,
\]
the probability distribution of $h$ is approximately given by 
\[
 P(h)dh\simeq \exp[-2.37-4.12\ln h-1.53 (\ln h)^{2}-0.13(\ln h)^{3}]dh.
\]
$P(h)$ takes a maximum at $h\simeq 0.178$, while 
${\sqrt{\langle h^{2} \rangle}}\simeq 0.419$.
Using 
\[
 P(s_{e}|\nu)ds_{e}=P(h)\frac{dh}{ds_{e}}ds_{e},
\]
we have
\[
 \sqrt{\langle s_{e}^2 \rangle} 
\simeq 5.96 \frac{\sqrt{1-\gamma^{2}}}{\gamma^{6}\nu}.
\]
Putting
$a =A_{{\rm ref}}s_{e}=Ch$, 
we have 
\[
 P_{a}(a )da =P(C^{-1}a )C^{-1}da .
\]

From the above argument and the equation 
\[
\sqrt{ \langle a_{*}^{2} \rangle}=A_{{\rm ref}}(\eta_{\rm ta})\sqrt{\langle s_{e}^{2} \rangle}, 
\]
we find the expression for the initial spin of PBHs for $\gamma\simeq 1$:
\begin{equation}
\sqrt{ \langle a_{*}^{2} \rangle}
\simeq \frac{5.96}{24\sqrt{3}\pi}x_{\rm ta}^{2}(1-f)^{-1/2}T_{v_{\rm CN}}(k_{0},\eta_{\rm ta})\sigma_{H}\sqrt{1-\gamma^{2}}\nu^{-1}. 
\label{eq:asigmanuinverse}
\end{equation}
Putting $x_{\rm ta}=2.14$, $T_{v_{\rm CN}}(k_{0},\eta_{\rm ta})=0.622$,
$\bar{\delta}_{H}=\bar{\nu}\sigma_{H}$, $\nu=(5/2)\bar{\nu}$
and $\bar{\delta}_{H}\simeq 0.768$, we find
\begin{equation}
\sqrt{\langle a_{*}^{2}\rangle} \simeq 
3.90\times 10^{-3}
(1-f)^{-1/2}\sqrt{1-\gamma^{2}}
\left(\frac{\nu}{8}\right)^{-2}
\label{eq:afinal}
\end{equation}
for the PBH mass
\begin{equation}
 M\simeq 1.14 (1-f)^{3/2}M_{H}.
\label{eq:mfinal}
\end{equation}
Eliminating $f$ from Eqs.~(\ref{eq:afinal}) and (\ref{eq:mfinal}), we obtain the
following simple expression
\begin{equation}
\sqrt{\langle a_{*}^{2}\rangle} \simeq 
4.01 \times 10^{-3}
\left(\frac{M}{M_{H}}\right)^{-1/3} 
\sqrt{1-\gamma^{2}}
\left(\frac{\nu}{8}\right)^{-2}.
\label{eq:amfinal}
\end{equation}

Although $f$ or $M$ is a free parameter in the present scheme, numerical
simulations strongly 
suggest $M\simeq M_{H}$ except for the near-critical case in which 
$M\ll M_{H}$~\citep{Musco:2012au,Escriva:2019nsa}. 
If we put $M=M_{H}$, $\gamma=0.85$, 
$\nu=8$, the above expression yields 
$\sqrt{\langle a_{*}^{2}\rangle }\simeq 
2.14\times 10^{-3}$.
Therefore, we conclude that $\sqrt{\langle a_{*}^{2}\rangle}=O(10^{-3})$ or even smaller for 
$M\simeq M_{H}$. 

Let us now discuss small PBHs formed in the near-critical case.
In this case, only the small fraction of PBHs are produced through the critical collapse, while the rest have $M\sim M_{H}$. Therefore, we should fix
$\nu$
at the scale of $M_{H}$.
Using Eq.~(\ref{eq:amfinal}), for example, we find $\sqrt{\langle
a^{2}_{*}\rangle }\simeq  2.14 \times 10^{-2}$ for $M=10^{-3}M_{H}$,
$\gamma=0.85$, and 
$\bar{\nu}=8$.
It also strongly suggests that the angular momentum will 
play an important role and may significantly suppress 
the formation of PBHs of 
$M\lesssim 10^{-8} M_{H}$, 
for which $\sqrt{\langle a_{*}^{2}\rangle}\agt 1 $. 

\subsection{Implications}

\begin{figure}[htbp]
\begin{center}
 \includegraphics[width=0.8\textwidth]{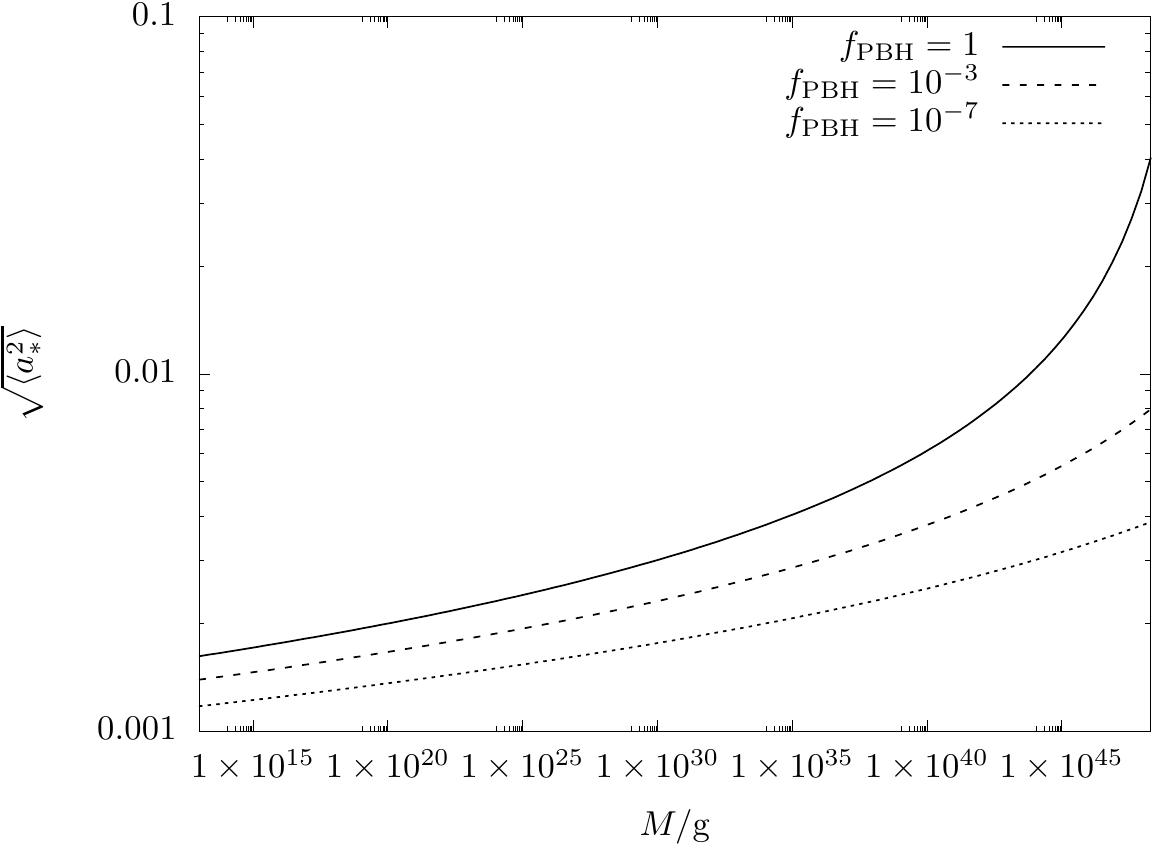}
\caption{\label{fg:afm} 
The standard deviation of the initial spins of the PBH, $\sqrt{\langle a_{*}^{2}\rangle }$, 
as a function of the PBH mass $M$ with fixed $f_{\rm PBH}$, where we have assumed that the PBH mass is equal to the horizon mass, i.e., $M=M_{H}$ and $\gamma=0.85$.
}
\end{center}
\end{figure}

Since our expression is given in terms of $\nu$,
the initial spin directly depends on the fraction $\beta_{0}(M_{H})$ of the universe which collapsed into black holes. If we use the Press-Schechter approximation 
as a rough estimate of $\beta_{0}(M_{H})$~\citep{Carr:1975qj}
\[
 \beta_{0}(M_{H})\simeq 
\sqrt{\frac{2}{\pi}}\frac{1}{\nu_{\rm th}}e^{-\nu_{\rm th}^{2}/2}
=\sqrt{\frac{2}{\pi}}\frac{2}{5}\frac{\sigma_{H}}{\bar{\delta}_{H,{\rm th}}}
\exp\left[-\left(\frac{5}{2}\right)^{2}\frac{\bar{\delta}_{H,{\rm th}}^{2}}{2\sigma_{H}^{2}}\right]
\]
we find a simple expression
\[
 \sqrt{\langle a_{*}^{2}\rangle} \simeq 
4.01\times 10^{-3}
\left(\frac{M}{M_{H}}\right)^{-1/3}
\sqrt{1-\gamma^{2}}\left[1-0.072\log_{10}
\left(\frac{\beta_{0}(M_{H})}{1.3\times 10^{-15}}\right)\right]^{-1},
\]
where 
$\nu$ 
is identified with 
$ {\nu}_{\rm th}$
and weak dependence on 
$\nu_{\rm th}$ 
in the logarithm is neglected.
For simplicity, let us concentrate on PBHs of $M\simeq M_{H}$.
Using the relation between $\beta_{0}(M)$ and $f_{\rm PBH}(M)$~\citep{Carr:2009jm}
\[
 \Omega_{\rm dm}f_{\rm PBH}(M)\simeq 10^{18}\beta_{0}(M)\left(\frac{M}{10^{15}\mbox{g}}\right)^{-1/2},
\]
we further obtain 
\begin{equation}
\sqrt{\langle a_{*}^{2}\rangle} 
\simeq 4.01 \times 10^{-3}
\frac{\sqrt{1-\gamma^{2}}}{1+0.036\left[21-2\log_{10}\left(\frac{f_{\rm PBH}(M)}{10^{-7}}\right)-\log_{10}
\left(\frac{M}{10^{15}\mbox{g}}\right)\right]}.
\label{eq:afm} 
\end{equation}
We plot Eq.~(\ref{eq:afm}) in Fig.~\ref{fg:afm}. In this figure, we can see that the larger $f_{\rm PBH}(M)$ and $M$, the larger $\sqrt{\langle a_{*}^{2}\rangle }$.
For example, $\sqrt{\langle a_{*}^{2}\rangle}$ of PBHs for 
$M=50M_{\odot}$ and $f_{\rm PBH}=1$ is about $ 3.3$ times larger than 
that for $M=10^{15}$ g and $f_{\rm PBH}=10^{-7}$.

\section{Summary and discussion}
We have applied \citet{Heavens:1988}'s
approach 
to the first-order effect on the spins of PBHs.
Although we have presented numerical values with two or three significant digits,
at present we admit that there is large uncertainty in 
modeling PBH formation. Nevertheless, we would like to claim that 
the standard deviation of the initial spins is 
given by $\sqrt{\langle a_{*}^{2}\rangle }=O(10^{-3})$ or even smaller 
for $M\simeq M_{H}$ based on peak theory.
We have obtained the expression
\begin{eqnarray*}
 \sqrt{\langle a_{*}^{2}\rangle} &\simeq& 
4.0\times 10^{-3}
\left(\frac{M}{M_{H}}\right)^{-1/3}
\sqrt{1-\gamma^{2}}
\left(\frac{\nu_{{\rm th}}}{8}\right)^{-2}
\nonumber \\
 &\simeq &
4.0\times 10^{-3}
\left(\frac{M}{M_{H}}\right)^{-1/3}
\sqrt{1-\gamma^{2}}\left[1-0.072\log_{10}\left(\frac{\beta_{0}(M_{H})}{1.3\times
					  10^{-15}}\right)\right]^{-1}.
\end{eqnarray*}
The above formula also implies that 
the higher the PBH formation probability $\beta_{0}(M_{H})$, the 
larger the standard deviation of the spins.
On the other hand, for PBHs of $M\ll M_{H}$ in the
near-critical case, we find $\sqrt{\langle a^{2}_{*}\rangle }$
can be much larger than those for PBHs of $M\simeq M_{H}$.

In comparison to the expression 
in \citet{DeLuca:2019buf},
the new estimate
has no overall factor $\Omega_{\rm dm}$, takes 
critical collapse into consideration, and gives an explicit expression
in terms of $\beta_{0}(M_{H})$. 
The proof of the nonexistence of the overall factor $\Omega_{\rm dm}$ 
is delegated to Appendix~\ref{sec:correction}.
From a physical point of view,
dark matter does not play any role in dynamics well in the radiation-dominated phase.
On the other hand, their assumption that the turn around occurs almost 
simultaneously to the horizon entry is supported if we take the horizon
entry of not the inversed wave number but the radius of the overdense region.
As for the gauge choice, we have taken the conformal Newtonian gauge to evaluate the angular momentum at the turn around. If we instead continued taking the CMC slicing as in \citet{DeLuca:2019buf}, the estimate would be further reduced by half approximately as seen in Fig.~\ref{fg:perturbations}.

Here, we would like to compare the present result with PBHs formed in the matter-dominated phase of the universe. 
As seen in this paper, in the radiation-dominated phase, since $\nu_{\rm th}$ is large suggested by the Jeans argument, 
peak theory implies that the region $\Sigma$ is nearly spherical and the
effect of tidal torque is suppressed. In fact, as we can see in
Eq.~(\ref{eq:se}), only the trace-free part $\mathcal{J}^{jl}$ of
$J^{jl}$ enters the expression of the angular momentum and we can see
that $\sqrt{\langle
(\mathcal{J}^{jl}\mathcal{J}_{jl})/(J^{jl}J_{jl})\rangle}=O(\nu^{-1}) $
for large $\nu$ and also $g(\eta)\sim k_{0}\sigma_{H}$.
In the matter-dominated phase, since $\nu_{\rm th}$ is vanishingly small in spherical symmetry, 
the region $\Sigma$ can be far from spherical and the tidal torque can
give the large amount of angular momentum on $\Sigma$.
Therefore, it is predicted that the angular momentum within the region
$\Sigma$ is so large
that PBH formation can be strongly suppressed and that 
the formed PBHs can have near-extremal spins~\citep{Harada:2017fjm}. 
This suggests that PBHs formed in the phase transition at which the
equation of state is softer than the radiation, their spins can be
larger than those formed in the radiation-dominated phase.

It would be interesting to remove the assumption of the narrow power
spectrum as the broad mass function of PBHs is intensively 
discussed from an observational point of
view~\citep{Carr:2016drx,Carr:2018poi}, although
the deviation of $\gamma$ from
unity might not change the result in 
orders of magnitude. Note also that although we have
investigated the first-order effect on the angular momentum, the
obtained result is apparently second order in terms of $\sigma_{H}$
as we can see in Eqs.~(\ref{eq:asigmanuinverse}) and (\ref{eq:afinal}), 
where $\nu^{-1}$ and $\bar{\nu}^{-1}$ are of the order of $\sigma_{H}$
because the threshold value of the PBH formation for 
the perturbation amplitude is of the order of unity. 
This means that the first-order
effect investigated here might be comparable to 
the second-order effect. In fact, 
in \citet{Mirbabayi:2019uph},
the second-order effect is estimated to 
$\sqrt{\langle a_{*}^{2}\rangle} \simeq \langle \zeta^{2}\rangle$.
Finally, it should be noted that
our analysis is based on the linear perturbation theory, which is 
not completely justified for perturbations that can generate PBHs.
In particular, the behaviors of the solutions at the final stage of 
black hole formation 
are highly nonlinear and cannot be predicted by linear
perturbation theory. In this line, the assumption of the conservation
of the nondimensional Kerr parameter after the decoupling from the
cosmological expansion should be confirmed by numerical simulations. 
It is clear that further investigations are necessary to answer the problem 
how large spins PBHs have.

\acknowledgements

T.H. was very grateful to B.~J.~Carr for indicating him this problem.
T.H. would also like to T.~Murata, K.~Nakashi, T.~Sato, and Y.~Watanabe 
for interesting discussion on peak theory.
The authors would like to thank K.~Nakao for his helpful comment and continuous 
encouragement. They would also like to thank the anonymous referee for the important 
comments and suggestions.
This work was supported by JSPS KAKENHI Grant
Numbers JP19H01895 (T.H. and C.Y.), JP19K03876 (T.H.), 
JP17H01131 (K.K.), 
and 
JP19J12007 (Y.K.)
and MEXT KAKENHI Grant Numbers JP19H05114 (K.K.) and
JP20H04750 (K.K.).
\appendix

\section{Peak theory} \label{sec:peaktheory}

We briefly review peak theory based on \citet{Heavens:1988}.
We treat the following fields as probability variables.
\begin{eqnarray*}
 \delta,~~
 \zeta_{i}=\frac{\partial \delta}{\partial x^{i}},~~
 \zeta_{ij}= \frac{\partial^{2}\delta}{\partial x_{i}\partial
  x_{j}},~~
 v^{i}_{~j}=\frac{\partial v^{i}}{\partial x^{j}}.
\end{eqnarray*}
The correlations of the above variables are given by
\begin{eqnarray}
 \langle \delta^{2} \rangle  &=& \sigma_{0}^{2}, ~~
 \langle \delta \zeta_{11}\rangle = -\langle \zeta_{1}\zeta_{1}\rangle
  =\cdots =-\frac{\sigma_{1}^{2}}{3}, ~~
 \langle \delta \tilde{v}_{11}\rangle = \cdots = -\frac{\sigma_{0}}{3}, \nonumber \\
 \langle \zeta_{11}^{2}\rangle &=& 3\langle \zeta_{11}\zeta_{22} \rangle
  =3\langle \zeta_{12}^{2}\rangle =\cdots
  =\frac{\sigma_{2}^{2}}{5}, \nonumber \\
 \langle \zeta_{11}\tilde{v}_{11}\rangle &=& 3\langle
  \zeta_{11}\tilde{v}_{22}\rangle =3\langle
  \zeta_{12}\tilde{v}_{12}\rangle =\cdots
  =\frac{\sigma_{1}^{2}}{5\sigma_{0}}, \nonumber \\
  \langle \tilde{v}_{11}^{2}\rangle &=& 3\langle
   \tilde{v}_{11}\tilde{v}_{22}\rangle =3\langle
   \tilde{v}_{12}^{2}\rangle =\cdots =\frac{1}{5},
\label{eq:variancetildevij}
\end{eqnarray}
and all other correlations vanish,
where 
\begin{eqnarray}
 \sigma_{j}^{2}&:=&
\int \frac{d^{3}{\bf k}}{(2\pi)^{3}}k^{2j}|\delta_{{\bf k}}|^{2} 
\label{eq:sigmaj}, \\
 \tilde{v}^{i}_{~j}&:=&
-\frac{1}{\sigma_{0}}
\int \frac{d^{3}{\bf k}}{(2\pi)^{3}}
\frac{k^{i}k_{j}}{k^{2}}\delta_{\bf  k} e^{i {\bf k}\cdot {\bf x}}.
\label{eq:vijtilde}
\end{eqnarray}
Putting the eigenvalues of $-\zeta_{ij}/\sigma_{2}$ as 
$\lambda_{1}$, $\lambda_{2}$, and 
$\lambda_{3}$ ($\lambda_{1}\ge \lambda_{2}\ge \lambda_{3}$), and 
\begin{eqnarray*}
 \nu&=& \delta/\sigma_{0}, ~~
 \xi_{1}= \lambda_{1}+\lambda_{2}+\lambda_{3}, ~~
 \xi_{2}=\frac{1}{2}(\lambda_{1}-\lambda_{3}), ~~
 \xi_{3}=\frac{1}{2}(\lambda_{1}-2\lambda_{2}+\lambda_{3}), \\
 w_{1}&=&\tilde{v}_{23}, ~~
 w_{2}=\tilde{v}_{13}, ~~
 w_{3}=\tilde{v}_{12},
\end{eqnarray*}
the probability distribution of $\nu$, ${\boldsymbol \lambda}$, and ${\bf w}$ at a peak is given by 
\[
 N_{pk}(\nu,\boldsymbol{\lambda},{\bf w})d\nu d^{3}{\bf
  \xi} d{\bf
  w}=\frac{B}{R_{*}^{3}}\exp(-Q_{4})F(\boldsymbol{\lambda})
d\nu  
d^{3}{\boldsymbol{\lambda}} d^{3}{\bf w},
\]
where 
\begin{eqnarray*}
B&=&\frac{3^{9/2}5^{4}}{2^{11/2}\pi^{9/2}(1-\gamma^{2})^{2}}, \\
2Q_{4}&=&\nu^{2}+\frac{(\xi_{1}-\gamma \nu)^{2}}{1-\gamma^{2}}+15\xi_{2}^{2}+5\xi_{3}^{2}+15\frac{w_{1}^{2}+w_{2}^{2}+w_{3}^{2}}{1-\gamma^{2}}, \\
F(\boldsymbol{\lambda})&=&\frac{27}{2}\lambda_{1}\lambda_{2}\lambda_{3}
(\lambda_{2}-\lambda_{3})(\lambda_{1}-\lambda_{3})(\lambda_{1}-\lambda_{2}), \\
R_{*}:&=& \sqrt{3}\frac{\sigma_{1}}{\sigma_{2}}, \quad \gamma:={\sigma_{1}^{2}}/({\sigma_{0}\sigma_{2}}), 
\end{eqnarray*}
We can see that the distribution of ${\bf w}$ is independent from other variables.

\section{Cosmological linear perturbations} 
\label{sec:perturbationtheory}

Here we briefly review the result of cosmological linear
perturbation theory that is necessary for the present paper. We basically follow the notation of \citet{Kodama:1985bj}.
We would like readers to refer to \citet{Kodama:1985bj}
or other reference for derivation. 
The scalar, vector and tensor harmonic functions $Y$, $Y_{i}$, $Y_{ij}$ in the flat space for scalar 
perturbations are defined as follows:
\[
 Y=Ce^{ik_{l}x^{l}}, ~~
 Y_{i}=-k^{-1}Y_{|i}, ~~
 Y_{ij}=k^{-2}\left(Y_{|ij}-\frac{1}{3}\delta_{ij}\Delta Y\right)=-\left(\frac{k_{i}k_{j}}{k^{2}}-\frac{1}{3}\delta_{ij}\right)Y, 
\]
where the roman indices are raised and lowered by $\delta^{ij}$ and
$\delta_{ij}$, respectively, and $Y$ satisfies
\[
 (\Delta +k^{2})Y=0. 
\]
The Fourier decomposition of the perturbations is given by  
\begin{eqnarray*}
 \delta(\eta,{\bf x})=\int \frac{d^{3}{\bf
  k}}{(2\pi)^{3}}\delta_{{\bf k}}(\eta)e^{i {\bf k}\cdot {\bf x}}, ~~
 \delta_{{\bf k}}(\eta)=\int d^{3}{\bf
  x}\delta(\eta,{\bf x})e^{-i {\bf k}\cdot {\bf x}},
\end{eqnarray*}
and so on. In the following in this section, we abbreviate $\delta_{{\bf k}}(\eta)$
as $\delta$ and so on. 
In Eq.~(\ref{eq:3+1}), 
we write the scalar perturbation of the metric tensor as follows:
\[
 \alpha=a(1+AY),~~
 \beta_{i}=-a^{2}BY_{i},~~
 \gamma_{ij}=\delta_{ij}+2H_{L}Y\delta_{ij}+2H_{T}Y_{ij}.
\]
The trace of the extrinsic curvature of the constant $\eta$ hypersurface is written as
\[
 K=K_{b}(1+{\cal K}_{g}Y).
\]
The perturbed quantities of the perfect fluid are written as 
\begin{equation}
 \rho=\rho_{b}(1+\delta Y), ~~
 p=p_{b}(1+\pi_{L}Y), ~~ 
 v^{i}=\frac{u^{i}}{u^{0}}=v Y^{i}. \label{eq:v}
\end{equation}
In the adiabatic process with $p=w\rho$ equation of state, we have
$ \pi_{L}=\delta$.
For the scalar perturbation, the infinitesimal coordinate transformation 
is given by 
\[
 \bar{\eta}=\eta+T(\eta)Y,~~\bar{x}^{j}=x^{j}+L(\eta)Y^{j}, 
\] 
where $T$ and $L$ are
arbitrary functions of $\eta$.
Under this coordinate transformation, the metric perturbation quantities transform as follows: 
\[
 \bar{A}=A-T'-{\cal H}T,~ 
 \bar{B}=B+L'+kT, ~
 \bar{H}_{L}= H_{L}-\frac{k}{n}L-{\cal H}T,~
 \bar{H}_{T}= H_{T}+kL.
\]
where $n$ is the dimension of the space,
$ {\cal H}:={a'}/{a}$, and the prime denotes the derivative with respect to $\eta$.
On the other hand, 
matter perturbation quantities transform as follows:
\[
 \bar{v}=v+L', ~~
 \bar{\delta}=\delta+n(1+w){\cal H}T,~~
 \bar{\pi}_{L}=\pi_{L}+3\frac{c_{s}^{2}}{w}(1+w){\cal H}T,
\]
where $c_{s}^{2}=w$ is the sound speed.
From the above, we can construct gauge-invariant perturbation quantities 
corresponding to $\delta$ and $V$ as follows:
\[
 \Delta=\delta+3(1+w){\cal H}k^{-1}(v-B),~~ V=v-k^{-1}H_{T}'.
\]

From the Einstein equation, we can derive the equations for the
gauge-invariant variables, $\Delta$ and $V$.
We present the solutions for the radiation-dominated phase of the universe
below: 
\begin{eqnarray*}
 \Delta(x)&=&D\sqrt{3}\left(\frac{\sin z}{z}-\cos z\right), \\
 V(x)&=&D\left[\frac{3}{4}\left(\frac{2}{z^{2}}-1\right)\sin
  z-\frac{3}{2}\frac{\cos z}{z}\right],
\end{eqnarray*}
where 
$D$ is an arbitrary constant, 
$z:=x/\sqrt{3}$, 
$x:=k\eta$, and 
a decaying mode is omitted.

In the CMC (${\cal K}_{g}=0$) slicing with $B=0$, 
using the above solutions for $\Delta(x)$ and $V(x)$, we can obtain
\begin{eqnarray}
\delta&=& D\frac{\sqrt{3}z^{2}}{z^{2}+2}\left(2\frac{\sin
						 z}{z}-\cos z\right), \label{eq:deltaCMC}\\
v&=&-\frac{3}{4}D\frac{(z^{2}-2)\sin z +2z\cos z}{z^{2}+2}, \label{eq:vcmc}\\
{\cal R}&=&D\frac{\sqrt{3}}{2}\frac{1}{z^{2}+2}\left(2\frac{\sin
						z}{z}-\cos z\right), 
\label{eq:calR}
\end{eqnarray}
where $ {\cal R}=H_{L}+\frac{1}{3}H_{T}$. In this gauge, 
$A$ and ${\cal R}$ are completely 
fixed, while $H_{T}$ and $H_{L}$ are fixed only up to a constant.

In the conformal Newtonian gauge, in which $H_{T}=B=0$, we can obtain
\begin{eqnarray}
 \delta &=&\sqrt{3}D\frac{2(z^{2}-1)\sin z+(2-z^{2})z\cos
  z}{z^{3}}, \\
 v &=&
\frac{3}{4}D\frac{(2-z^{2})\sin z-2z\cos z}{z^{2}} \label{eq:vCN}
, \\
 \Phi &=&-\frac{\sqrt{3}}{2}D\frac{\sin z-z\cos z}{z^{3}},
\label{eq:Phi}
\end{eqnarray}
where $H_{L}=-\Phi$ and $A=\Phi$. Thus, all perturbation quantities are completely fixed
in this gauge.

We can define transfer functions $T_{\delta_{\rm CMC}}$, $T_{v_{\rm CMC}}$, $T_{\delta_{\rm CN}}$, $T_{v_{\rm CN}}$ as follows:
\begin{eqnarray*}
\delta_{\rm CMC}(\eta)&=&T_{\delta_{\rm CMC}}(k,\eta)\Phi(0),~ 
v_{\rm CMC}(\eta)=T_{v_{\rm CMC}}(k,\eta)\Phi(0), \nonumber \\
\delta_{\rm CN}(\eta)&=&T_{\delta_{\rm CN}}(k,\eta)\Phi(0),~ 
v_{\rm CN}(\eta)=T_{v_{\rm CN}}(k,\eta)\Phi(0),
\end{eqnarray*}
where we can see $\Phi(0)=-D/(2\sqrt{3})=-(2/3){\cal R}(0)$ from
Eqs.~(\ref{eq:calR}) and (\ref{eq:Phi}). We should note that ${\cal R}(0)=-\zeta(0)$, where $\zeta$ is the curvature perturbation in
the uniform-density slicing.

\section{Nonexistence of the overall factor 
$\Omega_{\rm dm}$} \label{sec:correction}

Here, we show that the overall factor $\Omega_{\rm dm}$ in 
\citet{DeLuca:2019buf}'s expression 
should be removed. 
Although their notation is slightly different from ours, we consistently
continue to use our notation.

In the following, we follow the process of calculation in \citet{DeLuca:2019buf}.
They estimate the angular momentum at the horizon entry of the inversed wave
number saying that 
the turn around is just after the horizon entry.
Their analysis is confined to the CMC slicing, where 
$g(\tilde{\eta}_{H})=g_{{\rm CMC}}(\tilde{\eta}_{H})$ was
estimated to 
\begin{equation}
g_{{\rm CMC}}(\tilde{\eta}_{H})
\sim \left|\frac{T_{v_{\rm CMC}}(k_{0},\tilde{\eta}_{H})}{T_{\delta_{\rm CMC}}(k_{0},\tilde{\eta}_{H})}\right|k_{0}\tilde{\sigma}_{H},
\label{eq:gcmcetaHDeLuca}
\end{equation}
where $k_{0}$ is identified with $k_{H}$ in \citet{DeLuca:2019buf}, 
$\tilde{\eta}_{H}=k_{0}^{-1}$, and $\tilde{\sigma}_{H}$ is
$\sigma_{0}$ at $\eta=\tilde{\eta}_{H}$ without the long-wavelength limit.
Although our calculation does not reproduce their numerical value 
$|T_{v_{\rm CMC}}(k_{0},\tilde{\eta}_{H})/T_{\delta_{\rm
CMC}}(k_{0},\tilde{\eta}_{H})| \sim 0.5$ but 
gives a much smaller value $\simeq 0.0714$ at $x=1$,
this is not the origin of the factor $\Omega_{\rm dm}$.
Since ${\cal H}\propto a^{-1}$ in the radiation-dominated era
and ${\cal H}\propto a^{-1/2}$ in the matter-dominated era,
they probably inferred that 
\begin{equation}
 {\cal H}(\eta_{\rm eq})=\frac{{\cal H}_{0}}{(a(\eta_{\rm eq})/a_{0})^{1/2}},
\label{eq:DeLucawrong}
\end{equation}
where we have put $a(\eta_{0})=a_{0}$ and ${\cal H}(\eta_{0})={\cal H}_{0}$
and $\eta_{0}$ is the present conformal time.
This corresponds to Eq. (5.4) in \citet{DeLuca:2019buf}.
Then, using
\begin{eqnarray}
 \frac{a(\tilde{\eta}_{H})}{a_{0}}=\frac{a(\eta_{\rm eq})}{a_{0}}\left(\frac{{\cal H}(\eta_{\rm eq})}{{\cal
			    H}(\tilde{\eta}_{H})}\right)=\left(\frac{a(\eta_{\rm eq})}{a_{0}}\right)^{1/2}\left(\frac{{\cal H}_{0}}{{\cal
			    H}(\tilde{\eta}_{H})}\right)
\label{eq:aetaHa0wrong}
\end{eqnarray}
and defining $\tilde{M}_{H}$ as the mass within the Hubble horizon at
$\eta=\tilde{\eta}_{H}$, 
we find
\begin{equation}
 {\cal  H}(\tilde{\eta}_{H})=\frac{a(\tilde{\eta}_{H})}{2G\tilde{M}_{H}}=\left(\frac{a(\eta_{\rm eq})}{a_{0}}\right)^{1/2}\frac{{\cal
  H}_{0}}{{\cal H}(\tilde{\eta}_{H})}\frac{a_{0}}{2G\tilde{M}_{H}},
\label{eq:calHetaHwrong}
\end{equation}
and
\begin{equation}
 k_{0}=\left(
\frac{a(\eta_{\rm eq})}{a_{0}}\right)^{1/4}\sqrt{\frac{{\cal H}_{0}a_{0}}{2G\tilde{M}_{H}}}.
\label{eq:kHwrong}
\end{equation}
Moreover, using 
\begin{eqnarray*}
&& \rho_{b}(\tilde{\eta}_{H})\left(\frac{a(\tilde{\eta}_{H})}{a_{0}}\right)^{4}
\simeq
  \rho_{\rm rad}(\tilde{\eta}_{H})\left(\frac{a(\tilde{\eta}_{H})}{a_{0}}\right)^{4}
\simeq
  \rho_{\rm rad}(\eta_{0})
  =\Omega_{\rm dm}\frac{a(\eta_{\rm eq})}{a_{0}}\frac{3{\cal H}_{0}^{2}}{8\pi G a_{0}^{2}}, \\
&& k_{0}^{-1}={\cal H}^{-1}(\tilde{\eta}_{H})=\frac{2G\tilde{M}_{H}}{a(\tilde{\eta}_{H})},
\end{eqnarray*}
and identifying the mass of the PBH with $\tilde{M}_{H}$,
 they reached their conclusion
\begin{eqnarray}
 A_{{\rm ref},{\rm CMC}}(\tilde{\eta}_{H})
 \simeq \frac{\frac{4}{3} a_{0}^{4} \Omega_{\rm dm}\frac{a(\eta_{\rm eq})}{a_{0}}\frac{3{\cal H}_{0}^{2}}{8\pi
  Ga_{0}^{2}}\frac{1}{2}\left[\left(\frac{a(\eta_{\rm eq})}{a_{0}}\right)^{1/4}\sqrt{\frac{{\cal
		H}_{0}a_{0}}{2G\tilde{M}_{H}}}\right]^{-4}\tilde{\sigma}_{H}}{G\tilde{M}_{H}^{2}}
 \simeq \frac{\Omega_{\rm dm}}{\pi}\tilde{\sigma}_{H},
\label{eq:ArefetaHwrong}
\end{eqnarray}
where $g_{\rm CMC}(\tilde{\eta}_{H})\sim 0.5k_{0}\tilde{\sigma}_{H}$, $R_{*}\simeq \sqrt{3}k_{0}^{-1}$, and $1-f\simeq 1/3$ have been used.

In the following, we would like to redo the above estimate more carefully.
Let us keep $g_{\rm CMC}(\tilde{\eta}_{H})$ as in Eq.~(\ref{eq:gcmcetaHDeLuca})
and focus on the factor $\Omega_{\rm dm}$. 
Using $\tilde{M}_{H}=({4\pi}/{3})
(\rho_{b}a^{3})(\tilde{\eta}_{H})(k_{0}^{-1})^{3}$, we can directly get the following simple estimate:
\begin{eqnarray}
 A_{{\rm ref},{\rm
  CMC}}(\tilde{\eta}_{H})&=&\frac{\frac{4}{3}\left[a^{4}\rho_{b}g_{{\rm
					      CMC}}\right](\tilde{\eta}_{H})(1-f)^{5/2}R_{*}^{5}}
{G \tilde{M}_{H}^{2}}
\simeq 2 \left|\frac{T_{v_{\rm CMC}}(k_{0},\tilde{\eta}_{H})}{T_{\delta_{\rm CMC}}(k_{0},\tilde{\eta}_{H})}\right| \frac{\tilde{\sigma}_{H}}{\pi},
\label{eq:DeLucacorrect}
\end{eqnarray}
where we have used the Friedmann 
equation only at the formation of PBHs well in the radiation-dominated era.
We can see that there is no overall factor $\Omega_{\rm dm}$. 

Although the above derivation is complete, it might be useful to 
trace the calculation in \citet{DeLuca:2019buf} in a right way.
Assuming that the energy of the universe consists of 
radiation, dark matter, and the cosmological constant, 
the Friedmann equation implies
\begin{equation}
 H^{2}=H_{0}^{2}\left[\Omega_{\rm rad}\left(\frac{a_{0}}{a}\right)^{4}+\Omega_{\rm dm}\left(\frac{a_{0}}{a}\right)^{3}+\Omega_{\Lambda}\right],
\label{eq:Friedmanneq}
\end{equation}
Moreover, we assume that $\Omega_{\rm rad}\ll \Omega_{\rm dm}$, $\Omega_{\rm dm}\simeq 0.3$,
and $\Omega_{\Lambda}\simeq 0.7$. Then, we can safely neglect 
$\Omega_{\Lambda}$ at the matter-radiation equality $\eta=\eta_{\rm eq}$, when  
$\rho_{\rm rad}=\rho_{\rm dm}$. This immediately implies 
\[
 \frac{\Omega_{\rm rad}}{\Omega_{\rm dm}}=\frac{a(\eta_{\rm eq})}{a_{0}}. 
\]
Therefore, Eq.~(\ref{eq:Friedmanneq}) implies 
\[
{\cal H}(\eta_{\rm eq})=\frac{{\cal H}_{0}}{(a(\eta_{\rm eq})/a_{0})^{1/2}}
\sqrt{2\Omega_{\rm dm}}. 
\]
This corrects
Eq.~(\ref{eq:DeLucawrong}) or Eq. (5.4) in \citet{DeLuca:2019buf}.
This gives a factor $\sqrt{2\Omega_{\rm dm}}$ on the rightmost side of Eqs.~(\ref{eq:aetaHa0wrong}) and (\ref{eq:calHetaHwrong}) and a factor $(2\Omega_{\rm dm})^{1/4}$ on the rightmost side of Eq.~(\ref{eq:kHwrong}). Thus, there appears a 
factor $(2\Omega_{\rm dm})^{-1}$ on the rightmost side of Eq.~(\ref{eq:ArefetaHwrong})
and this $\Omega_{\rm dm}^{-1}$ cancels out the factor $\Omega_{\rm dm}$ from $\rho_{b}$. Then,
we reach the same expression as in Eq. (\ref{eq:DeLucacorrect}).


\begin{thebibliography}{}
%\cite{LIGOScientific:2018jsj}
\bibitem[Abbott et al.(2019)]{LIGOScientific:2018jsj}
Abbott, B.~P.,~\textit{et al.} [LIGO Scientific \& Virgo] 2019,
%``Binary Black Hole Population Properties Inferred from the First and Second Observing Runs of Advanced LIGO and Advanced Virgo,''
Astrophys. J. Lett. \textbf{882} no.2, L24
%doi:10.3847/2041-8213/ab3800
%[arXiv:1811.12940 [astro-ph.HE]].
%344 citations counted in INSPIRE as of 07 Sep 2020

%\cite{LIGOScientific:2020stg}
\bibitem[Abbott et al.(2020)]{LIGOScientific:2020stg}
Abbott, R., \textit{et al.} [LIGO Scientific \& Virgo] 2020,
%``GW190412: Observation of a Binary-Black-Hole Coalescence with Asymmetric Masses,''
Phys. Rev. D \textbf{102} no.4, 043015
%doi:10.1103/PhysRevD.102.043015
%[arXiv:2004.08342 [astro-ph.HE]].
%113 citations counted in INSPIRE as of 07 Sep 2020

%\cite{Arbey:2019jmj}
\bibitem[Arbey et al.(2020)]{Arbey:2019jmj}
Arbey, A., Auffinger,~J., \& Silk,~J. 2020,
%``Evolution of primordial black hole spin due to Hawking radiation,''
Mon. Not. Roy. Astron. Soc. \textbf{494} no.1, 1257-1262
%doi:10.1093/mnras/staa765
%[arXiv:1906.04196 [astro-ph.CO]].
%10 citations counted in INSPIRE as of 07 Sep 2020

%\cite{Bardeen:1985tr}
\bibitem[Bardeen et al.(1986)]{Bardeen:1985tr}
Bardeen, J.~M., Bond, J.~R., Kaiser,~N., \& Szalay, A.~S. 1986,
%``The Statistics of Peaks of Gaussian Random Fields,''
Astrophys. J. \textbf{304}, 15-61
%doi:10.1086/164143
%2432 citations counted in INSPIRE as of 26 Aug 2020

\bibitem[Bird et al.(2016)]{Bird:2016dcv}
  Bird, S., Cholis, I., Mu\~noz, J.~B., Ali-Ha\"imoud, Y., Kamionkowski, M.,
Kovetz, E.~D., Raccanelli, A., \& Riess, A.~G. 2016,
%``Did LIGO detect dark matter?,''
  Phys.\ Rev.\ Lett.\  {\bf 116}, no. 20, 201301 
%  doi:10.1103/PhysRevLett.116.201301
%  [arXiv:1603.00464 [astro-ph.CO]].
  %%CITATION = doi:10.1103/PhysRevLett.116.201301;%%

%\cite{Carr:1975qj}
\bibitem[Carr(1975)]{Carr:1975qj}
Carr, B.~J. 1975,
%``The Primordial black hole mass spectrum,''
Astrophys. J. \textbf{201}, 1-19
%doi:10.1086/153853
%671 citations counted in INSPIRE as of 26 Aug 2020

%\cite{Carr:2009jm}
\bibitem[Carr et al.(2010)]{Carr:2009jm}
Carr, B.~J., Kohri, K., Sendouda, Y., \& Yokoyama, J. 2010,
%``New cosmological constraints on primordial black holes,''
Phys. Rev. D \textbf{81}, 104019
%doi:10.1103/PhysRevD.81.104019
%[arXiv:0912.5297 [astro-ph.CO]].
%611 citations counted in INSPIRE as of 27 Aug 2020

\bibitem[Carr et al.(2020)]{Carr:2020gox}
Carr, B.~J., Kohri, K., Sendouda, Y., and Yokoyama, J.,
%``Constraints on Primordial Black Holes,''
[arXiv:2002.12778 [astro-ph.CO]].

%\cite{Carr:2018poi}
\bibitem[Carr \& K\"{u}hnel(2019)]{Carr:2018poi}
Carr, B.~J. \& K\"{u}hnel, F. 2019,
%``Primordial black holes with multimodal mass spectra,''
Phys. Rev. D \textbf{99} no.10, 103535
%doi:10.1103/PhysRevD.99.103535
%[arXiv:1811.06532 [astro-ph.CO]].
%12 citations counted in INSPIRE as of 26 Aug 2020

%\cite{Carr:2016drx}
\bibitem[Carr et al.(2016)]{Carr:2016drx}
Carr, B.~J., K\"{u}hnel, F., \& Sandstad, M. 2016,
%``Primordial Black Holes as Dark Matter,''
Phys. Rev. D \textbf{94} no.8, 083504
%doi:10.1103/PhysRevD.94.083504
%[arXiv:1607.06077 [astro-ph.CO]].
%521 citations counted in INSPIRE as of 26 Aug 2020
%\cite{Carr:2017jsz}
\bibitem[Carr et al.(2017)]{Carr:2017jsz}
Carr, B.~J., Raidal, M., Tenkanen, T., Vaskonen, V., \& Veerm\"{a}e, H. 2017,
%``Primordial black hole constraints for extended mass functions,''
Phys. Rev. D \textbf{96} no.2, 023514
%doi:10.1103/PhysRevD.96.023514
%[arXiv:1705.05567 [astro-ph.CO]].
%209 citations counted in INSPIRE as of 09 Sep 2020


%\cite{Chiba:2017rvs}
\bibitem[Chiba \& Yokoyama(2017)]{Chiba:2017rvs}
  Chiba, T. \& Yokoyama, S. 2017,
%  ``Spin Distribution of Primordial Black Holes,''
%  arXiv:1704.06573 [gr-qc].
Prog. Theor. Exp. Phys. {\bf 2017}, 083E01
  %%CITATION = ARXIV:1704.06573;%%
  %2 citations counted in INSPIRE as of 01 Jun 2017

\bibitem[Clesse \& Garc\'{i}a-Bellido(2016)]{Clesse:2016vqa}
  Clesse, S. \& Garc\'{i}a-Bellido, J. 2017,
%``The clustering of massive Primordial Black Holes as Dark Matter: measuring their mass distribution with Advanced LIGO,''
  Phys.\ Dark Univ.\  {\bf 15}, 142
%  doi:10.1016/j.dark.2016.10.002
%  [arXiv:1603.05234 [astro-ph.CO]].
  %%CITATION = doi:10.1016/j.dark.2016.10.002;%%

%\cite{Dasgupta:2019cae}
\bibitem[Dasgupta et al. (2020)]{Dasgupta:2019cae}
Dasgupta,~B., Laha~R. \& Ray~A. 2020,
%``Neutrino and positron constraints on spinning primordial black hole dark matter,''
Phys. Rev. Lett. \textbf{125} no.10, 101101
%doi:10.1103/PhysRevLett.125.101101
%[arXiv:1912.01014 [hep-ph]].
%41 citations counted in INSPIRE as of 11 Dec 2020

%\cite{DeLuca:2019buf}
\bibitem[De Luca et al.(2019)]{DeLuca:2019buf}
De Luca, V., Desjacques, V., Franciolini, G., Malhotra, A., \& Riotto, A. 2019,
%``The initial spin probability distribution of primordial black holes,''
JCAP \textbf{05}, 018
%doi:10.1088/1475-7516/2019/05/018
%[arXiv:1903.01179 [astro-ph.CO]].
%32 citations counted in INSPIRE as of 28 Jul 2020

%\cite{DeLuca:2020bjf}
\bibitem[De Luca et al.(2020)]{DeLuca:2020bjf}
De Luca, V., Franciolini, G., Pani, P., \& Riotto, A. 2020,
%``The evolution of primordial black holes and their final observable spins,''
JCAP \textbf{04}, 052
%doi:10.1088/1475-7516/2020/04/052
%[arXiv:2003.02778 [astro-ph.CO]].
%15 citations counted in INSPIRE as of 07 Sep 2020


%\cite{Escriva:2019nsa}
\bibitem[Escriv\`{a}(2020)]{Escriva:2019nsa}
Escriv\`{a}, A. 2020,
%``Simulation of primordial black hole formation using pseudo-spectral methods,''
Phys. Dark Univ. \textbf{27}, 100466
%doi:10.1016/j.dark.2020.100466
%[arXiv:1907.13065 [gr-qc]].
%14 citations counted in INSPIRE as of 05 Sep 2020


%\cite{Escriva:2019phb}
\bibitem[Escriv\`{a} et al.(2020)]{Escriva:2019phb}
Escriv\`{a}, A., Germani, C., \& Sheth, R.~K. 2020,
%``Universal threshold for primordial black hole formation,''
Phys. Rev. D \textbf{101} no.4, 044022
%doi:10.1103/PhysRevD.101.044022
%[arXiv:1907.13311 [gr-qc]].
%31 citations counted in INSPIRE as of 28 Sep 2020
 
%\cite{Germani:2018jgr}
\bibitem[Gernami \& Musco (2019)]{Germani:2018jgr}
Germani,~C. \& Musco,~I. 2019,
%``Abundance of Primordial Black Holes Depends on the Shape of the Inflationary Power Spectrum,''
Phys. Rev. Lett. \textbf{122} no.14, 141302
%doi:10.1103/PhysRevLett.122.141302
%[arXiv:1805.04087 [astro-ph.CO]].
%104 citations counted in INSPIRE as of 11 Dec 2020

%\cite{Harada:2013epa}
\bibitem[Harada et al.(2013)]{Harada:2013epa}
Harada, T., Yoo, C.~M., \& Kohri, K. 2013,
%``Threshold of primordial black hole formation,''
Phys. Rev. D \textbf{88} no.8, 084051
%doi:10.1103/PhysRevD.88.084051
%[arXiv:1309.4201 [astro-ph.CO]].
%144 citations counted in INSPIRE as of 28 Aug 2020

%\cite{Harada:2015yda}
\bibitem[Harada et al.(2015)]{Harada:2015yda}
Harada, T., Yoo, C.~M., Nakama, T., \& Koga, Y. 2015,
%``Cosmological long-wavelength solutions and primordial black hole formation,''
Phys. Rev. D \textbf{91} no.8, 084057
%doi:10.1103/PhysRevD.91.084057
%[arXiv:1503.03934 [gr-qc]].
%51 citations counted in INSPIRE as of 28 Aug 2020

%\cite{Harada:2017fjm}
\bibitem[Harada et al.(2017)]{Harada:2017fjm}
Harada, T., Yoo, C.~M., Kohri, K., \& Nakao, K.~I. 2017,
%``Spins of primordial black holes formed in the matter-dominated phase of the Universe,''
Phys. Rev. D \textbf{96} no.8, 083517
%doi:10.1103/PhysRevD.96.083517
%[arXiv:1707.03595 [gr-qc]].
%53 citations counted in INSPIRE as of 26 Aug 2020



%\cite{He:2019cdb}
\bibitem[He \& Suyama(2019)]{He:2019cdb}
He, M. \& Suyama, T. 2019,
%``Formation threshold of rotating primordial black holes,''
Phys. Rev. D \textbf{100} no.6, 063520
%doi:10.1103/PhysRevD.100.063520
%[arXiv:1906.10987 [astro-ph.CO]].
%11 citations counted in INSPIRE as of 07 Sep 2020

\bibitem[Heavens \& Peacock(1988)]{Heavens:1988}
Heavens, A. \& Peacock, J 1988., 
%``Tidal torques and local density maxima'', 
Mon. Not. R. astr. Soc. {\bf 232}, 339

%\cite{Kodama:1985bj}
\bibitem[Kodama \& Sasaki(1985)]{Kodama:1985bj}
Kodama, H. \& Sasaki, M. 1984,
%``Cosmological Perturbation Theory,''
Prog. Theor. Phys. Suppl. \textbf{78}, 1-166
%doi:10.1143/PTPS.78.1
%1330 citations counted in INSPIRE as of 26 Aug 2020

%\cite{Mirbabayi:2019uph}
\bibitem[Mirbabayi et al.(2020)]{Mirbabayi:2019uph}
Mirbabayi, M., Gruzinov, A., \& Nore\~{n}a, J. 2020,
%``Spin of Primordial Black Holes,''
JCAP \textbf{03}, 017
%doi:10.1088/1475-7516/2020/03/017
%[arXiv:1901.05963 [astro-ph.CO]].
%36 citations counted in INSPIRE as of 07 Sep 2020

%\cite{Musco:2018rwt}
\bibitem[Musco(2019)]{Musco:2018rwt}
Musco, I. 2019,
%``Threshold for primordial black holes: Dependence on the shape of the cosmological perturbations,''
Phys. Rev. D \textbf{100} no.12, 123524
%doi:10.1103/PhysRevD.100.123524
%[arXiv:1809.02127 [gr-qc]].
%51 citations counted in INSPIRE as of 09 Sep 2020

%\cite{Musco:2012au}
\bibitem[Musco \& Miller(2013)]{Musco:2012au}
Musco, I. \& Miller, J.~C. 2013,
%``Primordial black hole formation in the early universe: critical behaviour and self-similarity,''
Class. Quant. Grav. \textbf{30}, 145009
%doi:10.1088/0264-9381/30/14/145009
%[arXiv:1201.2379 [gr-qc]].
%109 citations counted in INSPIRE as of 26 Aug 2020

%\cite{Musco:2008hv}
\bibitem[Musco et al.(2009)]{Musco:2008hv}
Musco, I., Miller, J.~C., \& Polnarev, A.~G. 2009,
%``Primordial black hole formation in the radiative era: Investigation of the critical nature of the collapse,''
Class. Quant. Grav. \textbf{26}, 235001
%doi:10.1088/0264-9381/26/23/235001
%[arXiv:0811.1452 [gr-qc]].
%100 citations counted in INSPIRE as of 26 Aug 2020



%\cite{Nakamura:1997sm}
\bibitem[Nakamura et al.(1997)]{Nakamura:1997sm}
Nakamura, T., Sasaki, M., Tanaka, T., \& Thorne, K.~S. 1997,
%``Gravitational waves from coalescing black hole MACHO binaries,''
Astrophys. J. Lett. \textbf{487}, L139-L142
%doi:10.1086/310886
%[arXiv:astro-ph/9708060 [astro-ph]].
%208 citations counted in INSPIRE as of 24 Sep 2020

%\cite{Niemeyer:1999ak}
\bibitem[Niemeyer \& Jedamzik(1999)]{Niemeyer:1999ak}
Niemeyer, J.~C. \& Jedamzik, K. 1999,
%``Dynamics of primordial black hole formation,''
Phys. Rev. D \textbf{59}, 124013
%doi:10.1103/PhysRevD.59.124013
%[arXiv:astro-ph/9901292 [astro-ph]].
%190 citations counted in INSPIRE as of 09 Sep 2020

%\cite{Page:1976ki}
\bibitem[Page(1976)]{Page:1976ki}
Page, D.~N. 1976,
%``Particle Emission Rates from a Black Hole. 2. Massless Particles from a Rotating Hole,''
Phys. Rev. D \textbf{14}, 3260-3273
%doi:10.1103/PhysRevD.14.3260
%320 citations counted in INSPIRE as of 07 Sep 2020

%\cite{Polnarev:2006aa}
\bibitem[Polnarev \& Musco(2007)]{Polnarev:2006aa}
Polnarev, A.~G. \& Musco, I. 2007,
%``Curvature profiles as initial conditions for primordial black hole formation,''
Class. Quant. Grav. \textbf{24}, 1405-1432
%doi:10.1088/0264-9381/24/6/003
%[arXiv:gr-qc/0605122 [gr-qc]].
%68 citations counted in INSPIRE as of 26 Aug 2020

%\cite{Raidal:2017mfl}
\bibitem[Raidal et al.(2017)]{Raidal:2017mfl}
  Raidal, M., Vaskonen, V., \& Veerm\"{a}e, H. 2017,
%``Gravitational Waves from Primordial Black Hole Mergers,''
  JCAP {\bf 1709}, 037
%  doi:10.1088/1475-7516/2017/09/037
%  [arXiv:1707.01480 [astro-ph.CO]].
  %%CITATION = doi:10.1088/1475-7516/2017/09/037;%%
  %7 citations counted in INSPIRE as of 03 Oct 2017

%\cite{Sasaki:2016jop}
\bibitem[Sasaki et al.(2016)]{Sasaki:2016jop}
  Sasaki, M., Suyama, T., Tanaka, T., \& Yokoyama, S. 2016, 
%  ``Primordial black hole scenario for the gravitational wave event GW150914,''
Phys. Rev. Lett. {\bf 117} 061101 
%  arXiv:1603.08338 [astro-ph.CO].
  %%CITATION = ARXIV:1603.08338;%%
  %14 citations counted in INSPIRE as of 13 Jul 2016



%\cite{Shibata:1999zs}
\bibitem[Shibata \& Sasaki(1999)]{Shibata:1999zs}
Shibata, M. \& Sasaki, M. 1999,
%``Black hole formation in the Friedmann universe: Formulation and computation in numerical relativity,''
Phys. Rev. D \textbf{60}, 084002
%doi:10.1103/PhysRevD.60.084002
%[arXiv:gr-qc/9905064 [gr-qc]].
%177 citations counted in INSPIRE as of 09 Sep 2020

%\cite{Tokeshi:2020tjq}
\bibitem[Tokeshi et al. (2020)]{Tokeshi:2020tjq}
Tokeshi,~K., Inomata,~K. and Yokoyama~J. 2020,
%``Window function dependence of the novel mass function of primordial black holes,''
arXiv:2005.07153 [astro-ph.CO].
%5 citations counted in INSPIRE as of 11 Dec 2020

%\cite{Young:2019osy}
\bibitem[Young (2019)]{Young:2019osy}
Young,~S. 2019, 
%``The primordial black hole formation criterion re-examined: Parametrisation, timing and the choice of window function,''
Int. J. Mod. Phys. D \textbf{29} no.02, 2030002
%doi:10.1142/S0218271820300025
%[arXiv:1905.01230 [astro-ph.CO]].
%33 citations counted in INSPIRE as of 11 Dec 2020

%\cite{Yoo:2018kvb}
\bibitem[Yoo et al.(2018)]{Yoo:2018kvb}
Yoo, C.~M., Harada, T., Garriga, J., \& Kohri, K. 2018,
%``Primordial black hole abundance from random Gaussian curvature perturbations and a local density threshold,''
PTEP \textbf{2018} no.12, 123E01
%doi:10.1093/ptep/pty120
%[arXiv:1805.03946 [astro-ph.CO]].
%63 citations counted in INSPIRE as of 09 Sep 2020

%\cite{Yoo:2020dkz}
\bibitem[Yoo et al. (2020)]{Yoo:2020dkz}
Yoo, C.~M., Harada, T., Hirano, S., \& Kohri,~K. 2020,
%``Abundance of Primordial Black Holes in Peak Theory for an Arbitrary Power Spectrum,''
to appear in PTEP [arXiv:2008.02425 [astro-ph.CO]].
%1 citations counted in INSPIRE as of 26 Aug 2020

%\cite{Wald:1984rg}
\bibitem[Wald(1984)]{Wald:1984rg}
Wald, R.~M. 1984,
General Relativity, (Chicago, Chicago University Press)
doi:10.7208/chicago/9780226870373.001.0001
%643 citations counted in INSPIRE as of 26 Aug 2020

\end{thebibliography}
\end{document}